\documentclass[conf]{new-aiaa}
\usepackage[utf8]{inputenc}

\usepackage{graphicx}
\usepackage{subcaption} % Required for subfigure environment
\usepackage[margin=1in]{geometry}
\usepackage{caption}  
\usepackage{amsmath}
\usepackage[version=4]{mhchem}
\usepackage{siunitx}
\usepackage{longtable,tabularx}
\usepackage[table]{xcolor} 
\usepackage{booktabs}
\title{Public Service Algorithm: towards a transparent, explainable, and scalable content curation for news content based on editorial values}

\author{Ahmad Mel\footnote{Tech Lead, A European Perspective, mel@ebu.ch}}
\affil{AISBL EBU-UER, European Broadcasting Union, Brussels\\
AIDA, IDLab, Ghent University}
\author{Sebastien Noir\footnote{ Head of Software Engineering, Technology and Innovation, noir@ebu.ch }}
\affil{European Broadcasting Union, Geneva}

\begin{document}

\maketitle

\begin{abstract}

The proliferation of disinformation challenges traditional, unscalable editorial processes and existing automated systems that prioritize engagement over public service values. To address this, we introduce the Public Service Algorithm (PSA), a novel framework using Large Language Models (LLMs) for scalable, transparent content curation based on Public Service Media (PSM) inspired values. Utilizing a large multilingual news dataset from the 'A European Perspective' project, our experiment directly compared article ratings from a panel of experienced editors from various European PSMs, with those from several LLMs, focusing on four criteria: diversity, in-depth analysis, forward-looking, and cross-border relevance. Utilizing criterion-specific prompts, our results indicate a promising alignment between human editorial judgment and LLM assessments, demonstrating the potential of LLMs to automate value-driven curation at scale without sacrificing transparency. This research constitutes a first step towards a scalable framework for the automatic curation of trustworthy news content. 
\end{abstract}

\section{Introduction}

% The proliferation of disinformation, misinformation and propaganda across digital platforms presents an unprecedented challenge to the dissemination and discoverability of trustworthy information. Traditional methods for news content selection and placement are heavily reliant on human editorial judgment, a process that is inherently time-consuming, resource-intensive, and difficult to scale. In addition, it is perceived that end-users, news consumers, might not be familiar with the principles and rules governing the prioritization of the content, less so if this decisions are assisted by algorithms. And while automated news content curation systems have emerged, they often prioritize audience engagement metrics like click-through rates (CTR) and engagement, and fail to incorporate the crucial role of PSM inspired values into the selection and recommendation process. Furthermore, the lack of transparency in many automated systems raises concerns about potential biases, missing checks and balances, and the erosion of the editorial propositions.

The proliferation of disinformation, misinformation, and propaganda across digital platforms presents an ever growing challenge to the dissemination and discoverability of trustworthy information \citep{wardle2017information}. Traditional methods for news content selection and placement are heavily reliant on human editorial judgment, a process that is inherently time-consuming, resource-intensive, and difficult to scale \citep{napoli2015social}. In addition, it is perceived that end-users, news consumers, might not be familiar with the principles and rules governing the prioritization of the content—less so if these decisions are assisted by algorithms. And while automated news content curation systems have emerged, they often prioritize audience engagement metrics like click-through rates (CTR) and overall engagement, and fail to incorporate the crucial role of public service media (PSM)-inspired values into the selection and recommendation process \citep{helberger2019public}. Furthermore, the lack of transparency in many automated systems raises concerns about potential biases, missing checks and balances, and the erosion of the editorial propositions \citep{ananny2018seeing,bozdag2013bias}. To address these limitations, this study introduces the Public Service Algorithm (PSA), a novel framework leveraging the capabilities of Large Language Models (LLMs) to enable scalable, and interpretable content curation rooted in PSM-inspired editorial values.

Recent advances LLMs have catalyzed a growing body of research on using these systems to evaluate news content in terms of credibility, bias, and quality. Loru et al.~\cite{loru2025decoding} systematically assessed how state-of-the-art LLMs, such as GPT-4o and Gemini, judge the credibility and political bias of over 2,000 news outlets in a zero-shot setup. 
% Their findings indicate strong performance in detecting low-credibility sources but reveal inconsistency and political skew in evaluating mid-tier or ideologically charged outlets.  

Similarly, Pratelli et al.~\cite{pratelli2024evaluation} evaluated LLM judgments on individual news articles using six human-defined reliability criteria. They found substantial overlap between model and expert assessments, particularly in detecting sensational language.

Other recent work has emphasized the importance of interpretability and explainability in automated content assessment. The \textsc{Pastel} framework~\cite{leite2023pastel} prompted LLMs to identify 19 weakly supervised credibility signals such as bias or evidence presence from articles, which were then aggregated to classify veracity. This approach preserved explainability while achieving near-supervised performance, showcasing how LLMs can surface nuanced, interpretable indicators of trustworthiness. Additionally, Shailya et al.~\cite{lext2025trustworthy} introduced a benchmark to assess whether LLM-generated explanations are aligned with factual evidence, reinforcing the growing emphasis on trustworthy and transparent reasoning. 
% More broadly, \textsc{TrustLLM}~\cite{huang2024trustllm} benchmarked 16 models across eight trust dimensions, including fairness, robustness, and truthfulness, further highlighting the need for rigorous evaluation of LLMs not just on outcomes but on ethical and epistemic behavior.

In the context of news ecosystems, the literature has also explored how recommendation systems can amplify bias or undermine editorial values. Elahi et al.~\cite{elahi2021responsible} reviewed how algorithmic recommenders may inadvertently foster filter bubbles or promote misinformation and called for systems that center transparency, fairness, and accountability. Lu et al.~\cite{lu2020editorial} offered a practical pathway by integrating editorial values, such as diversity and serendipity, into recommender systems, achieving broader content exposure without compromising engagement. 

% On the user side, Mitova et al.~\cite{mitova2023transparency} found across five countries that people expect news recommender systems to be explainable and controllable, particularly when they are aware of underlying algorithms.

Our work builds on and extends this diverse literature,  by using LLMs to rate individual news articles and systematically comparing their assessments with human expert judgments. While prior research has explored source-level classification, weakly supervised veracity cues, or recommender-level transparency, we focus on article-level alignment, both in terms of rating accuracy and the interpretability of model rationales. By benchmarking model outputs against curated editorial standards, we examine how well LLMs align with expert expectations. 
The main goal of this approach, is to scale content selection with editorial input. 
This approach contributes to the broader goal of trustworthy AI in media by integrating rating performance, editorial values, and transparency into a unified evaluation framework, as a first step in a broader more comprehensive explainable and editorially-driven recommendation system. 

For this, we tap into the rich data resources present in "A European Perspective" (AEP) project, a large-scale collaborative news exchange between European Public Service organizations, that hosts millions of articles published in their original languages, and currently ingesting an average of two thousand new stories every day. Using news stories from AEP, we designed a preliminary study to first formalize the content selection process according to the defined values, rate those articles by experts and by LLMs, and then compare, analyze and interpret the results in a comprehensive statistical study. Also, we explore how LLMs can provide both a numerical rating for the content evaluated and some rationale on why the content is evaluated with a given score. In practice, the reasoning is performed before the numerical rating, as LLMs responses are generally better when prompted to reflect first, act later (chain of thought principle). This approach enhances interpretability: the reason for the rating can be both assessed and understood by journalists offering the opportunity to enhance the editorial processes.
The overarching goal is to identify and promote articles characterized by their diversity, in-depth quality, forward-looking perspectives, and cross-border relevance. The first step of this study focuses on scaling the automation of the rating process against those dimensions using LLMs. Incoming articles to the AEP database will be automatically rated by PSA according to the aforementioned criteria, which will be used by editors as threshold for content selection to users. 

Our contributions in this research are:
\begin{enumerate}
\item The design and implementation of the Public Service Algorithm (PSA). This framework uses LLMs to rate news articles against four specific criteria, directly addressing the need for automated and scalable application of editorial judgment for high-value content identification.
\item The development of a human editorial baseline. We collected ratings from 30 experienced editors on a set of news articles. We then performed inter-rater reliability analysis on these human judgments to quantify their agreement and provide an empirical ground truth.
\item A direct comparison of LLM ratings with human editorial judgments. Our study demonstrates the capacity of LLMs to replicate human editorial decisions similarly to~\cite{pratelli2024evaluation} , particularly in ranking articles based on PSM values. The results indicate LLMs can align with expert consensus in selecting content.
\end{enumerate}

In our experimental setup, we selected a representative dataset of news articles for both human and LLM evaluations. The first phase involved a comprehensive human evaluation. Thirty experienced editors from participating AEP organizations independently rated various news stories based on the aforementioned key criteria. Each editor assigned a score from 1 to 5 on an incremental scale for each criterion, resulting in over 4000 individual ratings. Subsequently, the rating was followed by an analysis focused on inter-rater reliability, the identification of potential biases, and the detection of outliers in human judgment, given the subjective nature of the task. The second phase focused on evaluating the performance of several state-of-the-art commercial and open-weight LLMs (including OpenAI’s GPT-4, Meta’s Llama, among others) using custom prompts to elicit assessments aligned with each of the PSM criteria, applying these prompts to the same selection of articles evaluated by the editors. This was followed by a rigorous statistical analysis to compare the LLM-generated ratings with the human ratings, exploring the degree of correlation and alignment between the two. In addition to comparing different LLMs' performance, we investigated the internal consistency of the ratings provided by each LLM, given their inherent non-deterministic nature. 
% The analyses included inter-rater reliability calculations (adapted for the LLM context), correlation analysis, and outlier detection.
% , as well as ANOVA to analyze variations in ratings based on several article characteristics including: topic, category, source organization, language, and length.

Finally we discuss the results of the analysis and the limitations of our experimental setup and we highlight how can future research address these by exploring the integration of user engagement metrics to assess the real-world impact of content selected based on the PSA evaluations, and investigating the performance of fine-tuned LLMs specifically trained to recognize and prioritize these PSM inspired values.

This research constitutes a significant initial step towards achieving a value-driven, interpretable, and automation-assisted content curation framework to showcase more trusted and valuable news content.

\section{Methodology And Experimental Setup}

In this section we present our methodology and experimental setup, including the  dataset, evaluation criteria used for rating, the different LLMs used, and the evaluation metric for agreement analysis. 

This study employs a comparative experimental design to explore the use of LLMs for automating the rating of news articles against predefined editorial values. The primary objective is to determine the degree of alignment between ratings generated by LLMs and those provided by human experts, thereby assessing the feasibility of augmenting manual curation processes.

\begin{figure}[htbp]
    \centering
    \captionsetup{justification=centering} % Centers the caption text

    % Subfigure 1: Box Plot showing overall variability or per-criterion variability
        \includegraphics[width=\linewidth]{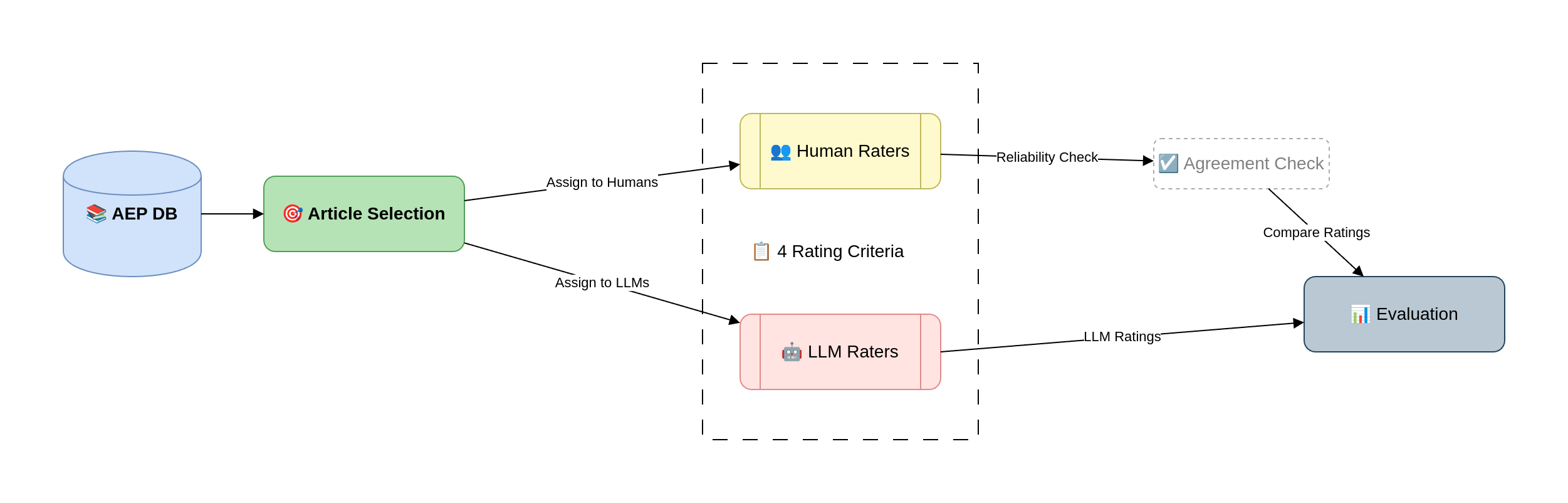} 

        \caption{Overview of the experimental setup for PSA. A representative sample taken from the AEP database, to be rated independently by editors and LLMs, on four different criteria and then assessed to compare the agreement performance between the LLMs and editors.} 
                    \label{fig:psa_setup}

\end{figure}

The procedure was designed to facilitate a direct and rigorous comparison between human editorial judgment and automated LLM performance, adhering to the same core methodology: first, the editorial team refined the description of the evaluation criteria to ensure clarity for all participants and LLMs. Following this, a formal briefing session was conducted with the participating editors from the  "A European Perspective" project to align their understanding of the goals, rating scales, and procedural guidelines. A sample of articles is selected from AEP, and data collection then proceeded in parallel: editors submitted their ratings, while the selected LLMs concurrently evaluated the same set of articles. The process is summarized in figure~\ref{fig:psa_setup}. 

By directly comparing the output ratings on those criteria, we aim to measure the reliability, consistency, and potential of LLMs as a scalable rating tool.

\subsection{Dataset and Article Selection}
To ensure a representative sample, a dataset of 30 news articles were randomly selected from the content repository of the AEP project\footnote{The dataset is available on this url: \url{https://overlay.yepnews.eu/assets/psa/dataset/psa_data.zip}}. This selection encompassed articles from various participating public service media organizations, covering a range of topics and original languages. All the languages have been translated into english and verified by human. This approach was intended to mitigate selection bias and test the robustness of both human and machine rating systems across diverse content.

\subsection{Selected LLMs}
To evaluate article-level on the different criteria, we employ a diverse set of state-of-the-art~\footnote{at the time}  LLMs, representing a range of architectures, sizes, and licenses. Our model suite includes both proprietary and open-weight models:

\begin{itemize}
    \item \textbf{GPT-4o}~\cite{openai_gpt4o_2024}, OpenAI’s multimodal, omni-directional model capable of reasoning over text, images, and audio, with strong zero-shot capabilities.
    \item \textbf{LLaMA~3.1 (405B)} and \textbf{LLaMA~3 (70B)}~\cite{meta_llama3_2024}, Meta's latest dense decoder models offering long-context understanding and strong performance across language tasks.
    \item \textbf{Mistral Large}~\cite{mistral_large_2023}, a performant instruction-tuned model optimized for general-purpose reasoning.
    \item \textbf{Mistral NeMo}, a smaller resource-efficient variant.
    \item \textbf{Command R+ (08/2024)}~\cite{cohere_command_r_plus_2024}, a long-context model by Cohere optimized for retrieval-augmented generation and robust instruction following.
    \item \textbf{Qwen-25.72B Instruct}~\cite{qwen_25_72b}, an instruction-tuned Chinese-English model with strong multilingual comprehension.
    \item \textbf{WizardLM-2 8$\times$22B}~\cite{wang_wizardlm_2023}, an ensemble of fine-tuned models trained with synthetic instructions.
\end{itemize}

This collection allows us to examine how different LLMs rate news articles in zero-shot settings, and how closely their judgments align with human expert benchmarks across multiple dimensions of credibility.

\subsection{Criteria}
The four editorial criteria were developed and refined by the PSA steering committee, composed primarily of senior editors from various public media organizations participating in the AEP project. First developed in 2022, these criteria are grounded in European public service values and were gradually adapted to reflect the core principles of the AEP initiative. Their primary purpose is to support editors in curating content for AEP in alignment with shared editorial goals.

To ensure usability in fast-paced newsroom environments, the definitions were intentionally kept concise, only as long as necessary, recognizing that experienced journalists, familiar with their own editorial standards, would be capable of applying them with professional judgment and discretion. In contrast, when designing prompts for LLMs, the criteria had to be translated into more explicit, detailed formulations. Given that LLMs can process long texts rapidly but lack the contextual training of human editors, the prompts were extended to include elaborations and specific examples that extend the original definitions. More importantly, translating human descriptions to LLMs instructions remains a challenge worthy of exploring in future research. 

To preserve transparency and editorial integrity, care was taken to ensure that these LLM prompts were faithful extensions of the original criteria and not reinterpretations.  Below is a summary of the criteria used, a more detailed version is present in appendix~\ref{annex:psa_crit}

\begin{itemize}
   \item In-depth analysis: Stories that orientate audiences about important issues or events by providing expert analysis 
based on facts

   \item Diversity: Stories about under-reported communities that add new voices and perspectives. 

   \item Cross-border relevance: Stories that have a universal aspect resonating with a broader European audience. 

   \item Forward looking: Stories showcasing innovative, forward-looking and constructive approaches to universal 
problems.

\end{itemize}

% A follow-up session with the editors to discuss the variations in the ratings and evaluate the differences in their ratings. 

\subsection{Human Expert Evaluation}
\label{sec:editors}
Initially, a panel of 30 experienced editors from AEP partner organizations was recruited to provide the human rating baseline. Editors were invited to independently rate the 30 selected articles against the four defined criteria, and their active contribution resulted in a variable number of ratings per article. To standardize the assessment and facilitate inter-rater reliability, participants were provided with an improved editorial guide that clearly defined each criterion and the rating scale. The goal of this phase was to generate a set of expert-driven scores to serve as the benchmark for evaluating LLM performance. 
Note, while practically speaking, of the 30 editors, 27 participated, and fewer rated all articles on all criteria. This provided a challenge for a comprehensive evaluation. This is further discussed in section~\ref{sec:lim}.

\subsection{Automated LLM Evaluation}
The same set of articles was evaluated using eight different LLMs. For the automated rating, a series of experiments was conducted where the LLMs were prompted to rate each article according to the four specified criteria. Each criterion had a dedicated, custom-engineered prompt designed to elicit a numerical score reflecting the article's alignment with that value (details present in appendix~\ref{annex:psa_crit}. This direct, criterion-by-criterion approach allowed for a granular comparison between the LLM-generated ratings and the human expert baseline. Additionally, the LLMs were prompted to provide a rationale for their ratings, this would be used to further improve the prompts in subsequent trials.
The selection of multiple LLMs, was intended to compare performance and identify the most suitable model for scalable deployment.

\subsection{Inter-Rater Reliability}
\label{sec:icc}
To evaluate the agreement between human raters on article quality assessments, we use the \textit{Intraclass Correlation Coefficient (ICC)}, a statistical measure commonly employed for assessing the reliability of ratings when multiple raters score the same items. Specifically, we use the two-way random-effects model with absolute agreement and average measures, denoted as \textbf{ICC(2,k)}~\cite{shrout1979icc}.

The ICC(2,k) answers the question: \textit{To what extent can the average score across $k$ human raters be trusted as a reliable representation of the true quality of each article?}

This formulation assumes that:
\begin{itemize}
    \item Each article is rated by the same $k$ raters.
    \item The raters are a random sample from a larger population of possible evaluators.
    \item We are interested in the absolute agreement among raters, not just consistency.
\end{itemize}

As such, ICC is appropriate when the average of multiple raters is used to produce the final score,  which is common in content annotation and quality evaluation tasks.
Note that, while ICC assumptions include random sampling, the study employs a purposive sample of relevant domain experts, which might affect statistical generalizability. 
\subsubsection*{Model Description}

The ICC(2,k) is computed using a two-way random-effects ANOVA model. The formula is:

\begin{equation}
\text{ICC(2,k)} = \frac{MS_B - MS_E}{MS_B + \frac{MS_R - MS_E}{n} + \frac{k-1}{n} MS_E}
\end{equation}

where:
\begin{itemize}
    \item $MS_B$: Mean square between targets (e.g., between articles)
    \item $MS_R$: Mean square between raters
    \item $MS_E$: Mean square error (residual)
    \item $n$: Number of targets
    \item $k$: Number of raters
\end{itemize}

\subsubsection*{Interpretation}
This model and its interpretation follow established guidelines for intraclass correlation analysis showcased by Koo and Li~\cite{koo2016guideline}.

ICC values range from 0 to 1, where:
\begin{description}
    \item[] \textbf{< 0.5}: Poor reliability
    \item[] \textbf{0.5 – 0.75}: Moderate reliability
    \item[] \textbf{0.75 – 0.9}: Good reliability
    \item[] \textbf{> 0.9}: Excellent reliability
\end{description}

ICC(2,k) value reflects how much of the total variance in the ratings is attributable to actual differences between the items, rather than inconsistencies across raters. The higher the ICC(2,k), the more trustworthy the mean ratings are as stable measurements of article quality. We computed ICC(2,k) using the \texttt{pingouin} package in Python.
Note that, while ICC requires a full matrix of targets and raters, as mentioned in section~\ref{sec:editors}, not all editors rated all the articles on all criteria. This poses a challenges for ICC application, thus we applied a pruning approach where we ignored raters that did not complete all ratings on a minimal set of articles per criteria. Given that ICC was applied per criteria, this minimized the elimination needed. This is further discussed in section~\ref{sec:lim}.

\subsection*{Evaluation Metrics}

To evaluate how well the article rankings generated by LLMs align with human editorial judgment, we adopt two complementary metrics: Normalized Discounted Cumulative Gain (NDCG@5) and Precision@5. These are computed over the top-5 ranked articles selected by each model from a shared pool of 30 articles.
\subsubsection*{Normalized Discounted Cumulative Gain (NDCG@5)}

NDCG@5~\cite{jarvelin2002cumulated} measures the quality of the model’s top-ranked outputs by taking both relevance and rank position into account. In our setup, relevance is derived from real-valued scores assigned by human editors, capturing nuanced aspects of editorial criteria.

The Discounted Cumulative Gain (DCG) is computed as:

\[
\mathrm{DCG@5} = \sum_{i=1}^{5} \frac{\text{rel}_i}{\log_2(i + 1)}
\]

where $\text{rel}_i$ is the human-assigned relevance score for the article at rank $i$. To normalize this score, the Ideal DCG (IDCG) is calculated by sorting the same articles by human scores:

\[
\mathrm{NDCG@5} = \frac{\mathrm{DCG@5}}{\mathrm{IDCG@5}}
\]

The logarithmic discount favors placing more relevant articles near the top. NDCG@5 ranges from 0 (poor ranking) to 1 (perfect alignment), and supports graded relevance judgments. Higher values indicate that a model’s ranking agrees more closely with human preferences.

\subsubsection*{Precision@5}

To complement NDCG@5, we also compute Precision@5, which provides a simpler, rank-insensitive measure of agreement. This metric captures how many of the articles selected in the model’s top-5 also appear in the human top-5 set. It is defined as:

\[
\text{Precision@5} = \frac{|\text{Top-5}_{\text{model}} \cap \text{Top-5}_{\text{human}}|}{5}
\]

Precision@5 ranges from 0 to 1, where 1 indicates full agreement in top-5 selections. Unlike NDCG, it treats all matching articles equally regardless of their specific position in the ranking, making it useful for evaluating overlap without requiring fine-grained order alignment.

Note that,  the human-assigned relevance score is an aggregate derived from judgments with quantified variability, thus the alignment measured by NDCG and Precision reflects an alignment with this aggregated, variable human consensus, rather than a perfectly stable and universally agreed-upon truth.

\section{Results and discussions}
This section presents the empirical findings regarding human rater agreement, the comparative rating distributions of LLMs and human editors, and a detailed analysis of LLM performance against human-derived benchmarks.

\subsection{Analysis of Human Rater Agreement}

\begin{figure}[htbp]
    \centering
    \includegraphics[width=0.9\textwidth]{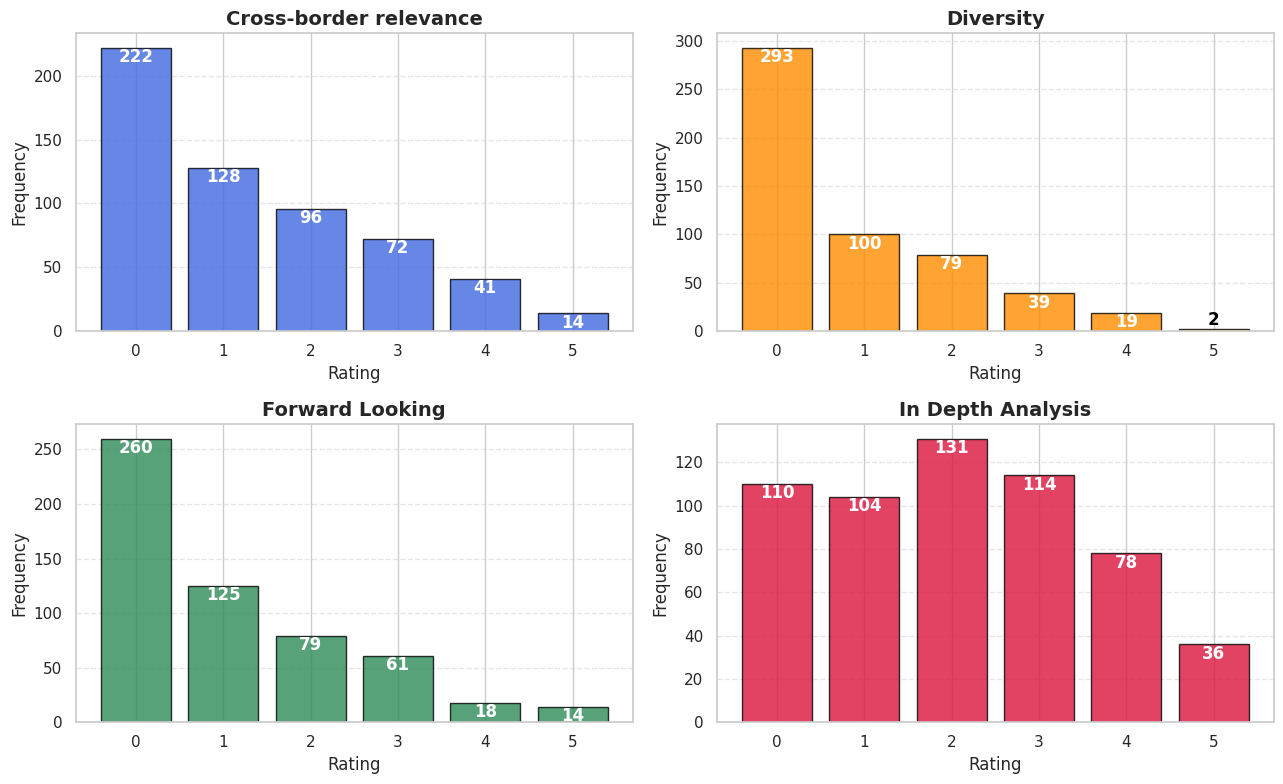}
    \captionsetup{justification=centering}
    \caption{The frequency histogram of human ratings per criteria.}
    \label{fig:human_hist}
\end{figure}

Before analyzing the agreement between the editors, we showcase the frequency histogram of human ratings per criteria in figure~\ref{fig:human_hist}. This highlights the prevalent assignment of lower scores (0.0 to 1.0) for 'Cross-border relevance', 'Diversity', and 'Forward Looking' criteria. For instance, 'Diversity' received a score of 0.0 in 293 instances, representing the most frequent rating for this criterion. In contrast, 'In-Depth Analysis' exhibits a more balanced distribution across the rating scale, indicating a wider range of perceived analytical depth among articles.

\begin{figure}[htbp]
    \centering
    \includegraphics[width=0.9\textwidth]{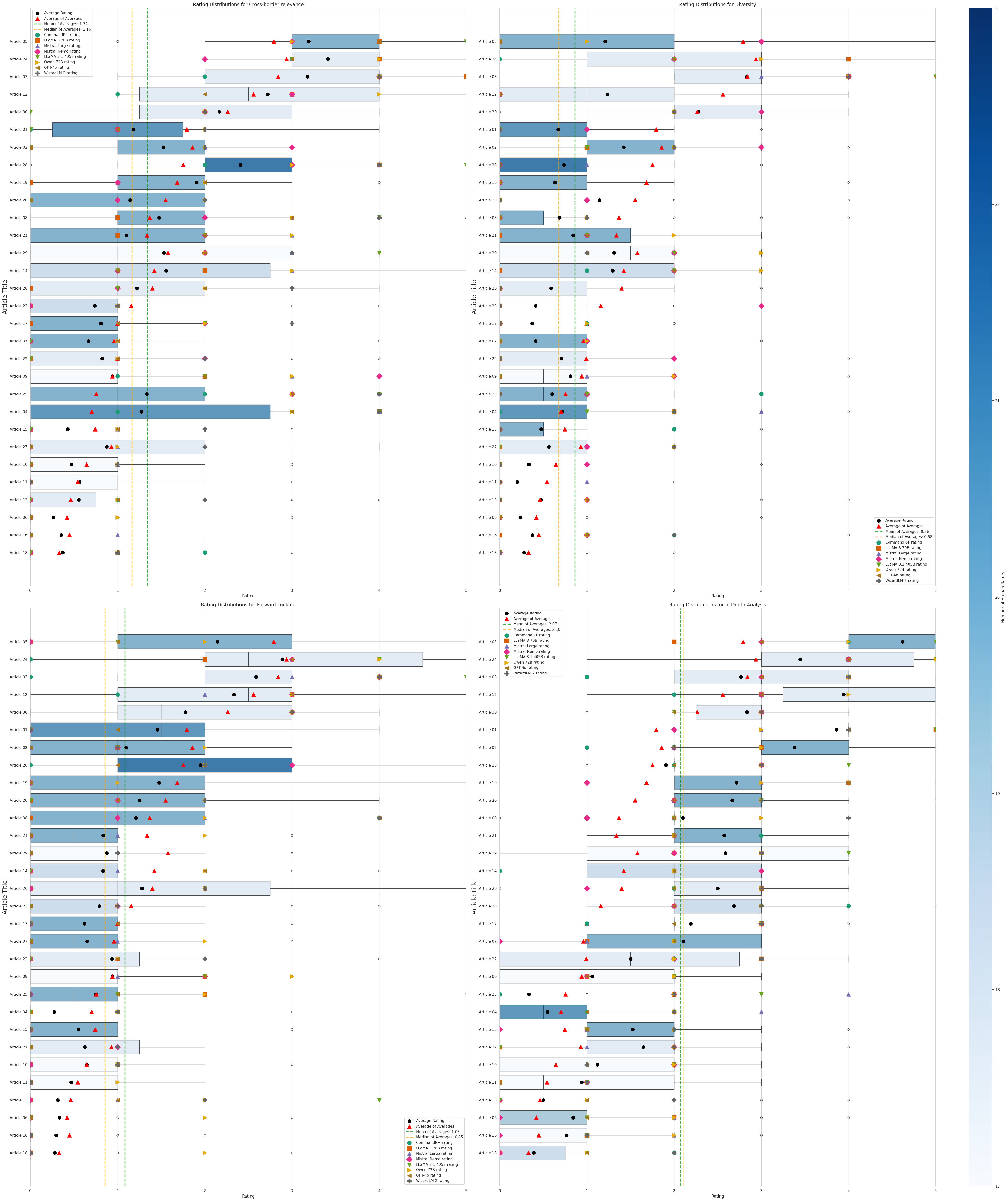}
    \captionsetup{justification=centering}
    \caption{Box plot showing inter-rater variability across rating criteria. Each box represents the distribution of ratings for a single criterion across multiple human raters.}
    \label{fig:rater_agreement_visuals}
\end{figure}

A principal finding from the human evaluation phase is the observed inter-rater variability across most articles and criteria. Visual inspection of the box plots in figure~\ref{fig:rater_agreement_visuals} reveals good consensus, particularly for articles that were rated low (e.g., scoring 0.0) where divergence was less pronounced. The box plots exhibit variable interquartile ranges (IQR) and an average minimum/maximum spread of approximately 2 points, indicating that editors often assigned a broad spectrum of scores to the same article for certain criteria. For instance, this variability is notably apparent for 'Diversity' and 'Forward-Looking', whereas 'In-Depth Analysis' and 'Cross-border Relevance' demonstrate tighter distributions. The presence of numerous outliers further highlights instances where individual judgments departed significantly from the central tendency of the group.

\begin{table}[htbp]
\centering
\caption{Inter-Rater Reliability Scores (ICC(2,k)) by Criterion}
\label{tab:icc_scores}
\begin{tabular}{l r}
\toprule
\textbf{Criterion} & \textbf{ICC Score} \\
\midrule
Cross-border relevance & 0.891 \\
Diversity              & 0.791 \\
Forward Looking        & 0.747 \\
In Depth Analysis      & 0.942 \\
\bottomrule
\end{tabular}
\end{table}

To formally quantify this level of agreement, we employed ICC(2,k), a robust statistical measure for inter-rater reliability. This descriptive statistic assesses the consistency or conformity of ratings by quantifying how much of the total variance in the data is attributable to differences among articles versus variance from random error or variations between raters. A higher score indicates better reliability (see section~\ref{sec:icc}). Although the box plots highlight individual judgment variations, the ICC scores, summarized in Table~\ref{tab:icc_scores}, indicate that the human ratings demonstrate a good overall ability to differentiate between articles based on their actual differences rather than inconsistencies in rater application of criteria. The higher the ICC, the more reliably the average rating reflects the distinct quality of an article for that criterion.

Given that the main goal is to use the results of the human rating to guide and assess the performance of the LLMs, the ICC(2,k) is considered a valuable approach.

\subsection{Rating distribution}

\begin{figure}[htbp]
    \centering
    % --- Row 1 ---
    \begin{subfigure}[t]{0.47\textwidth}
        \centering
        \includegraphics[width=\linewidth]{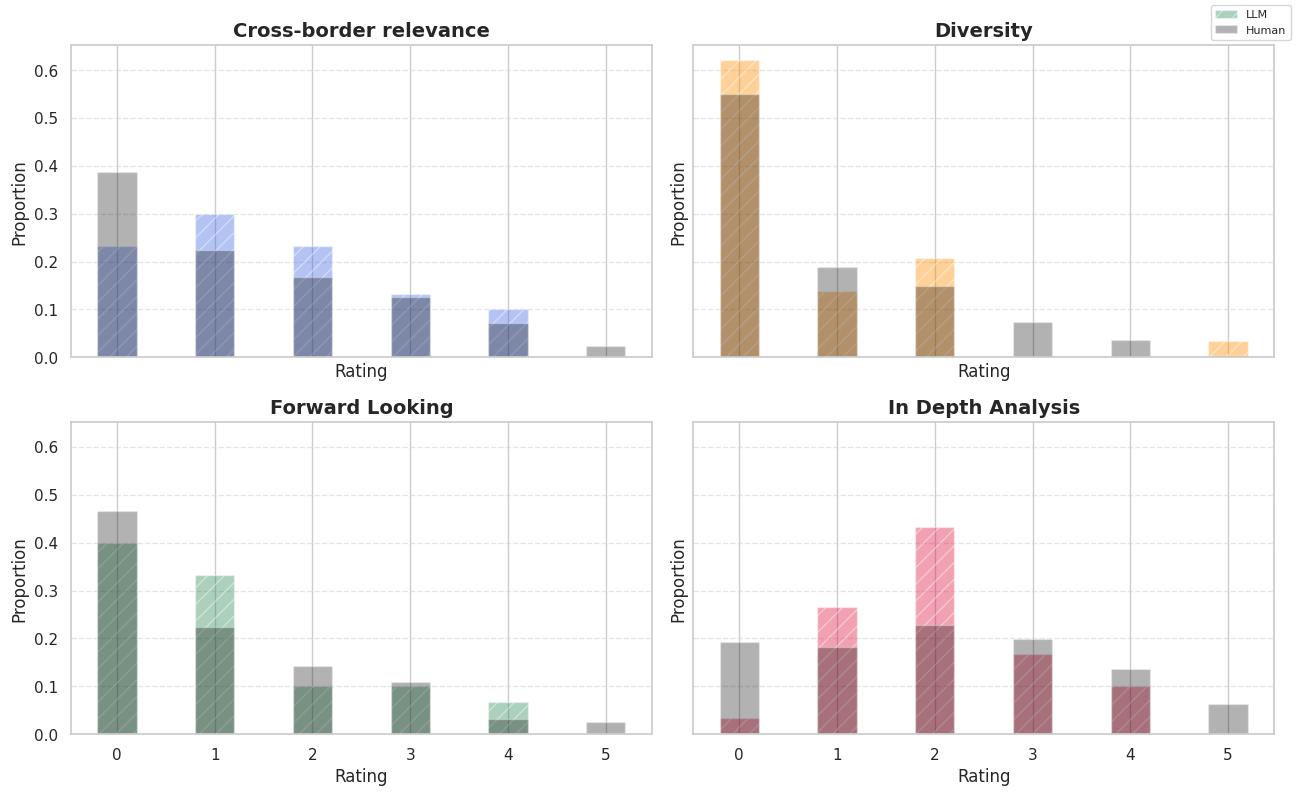}
        \caption{GPT-4o}
        \label{fig:dist_gpt4o}
    \end{subfigure}
    \hfill
    \begin{subfigure}[t]{0.47\textwidth}
        \centering
        \includegraphics[width=\linewidth]{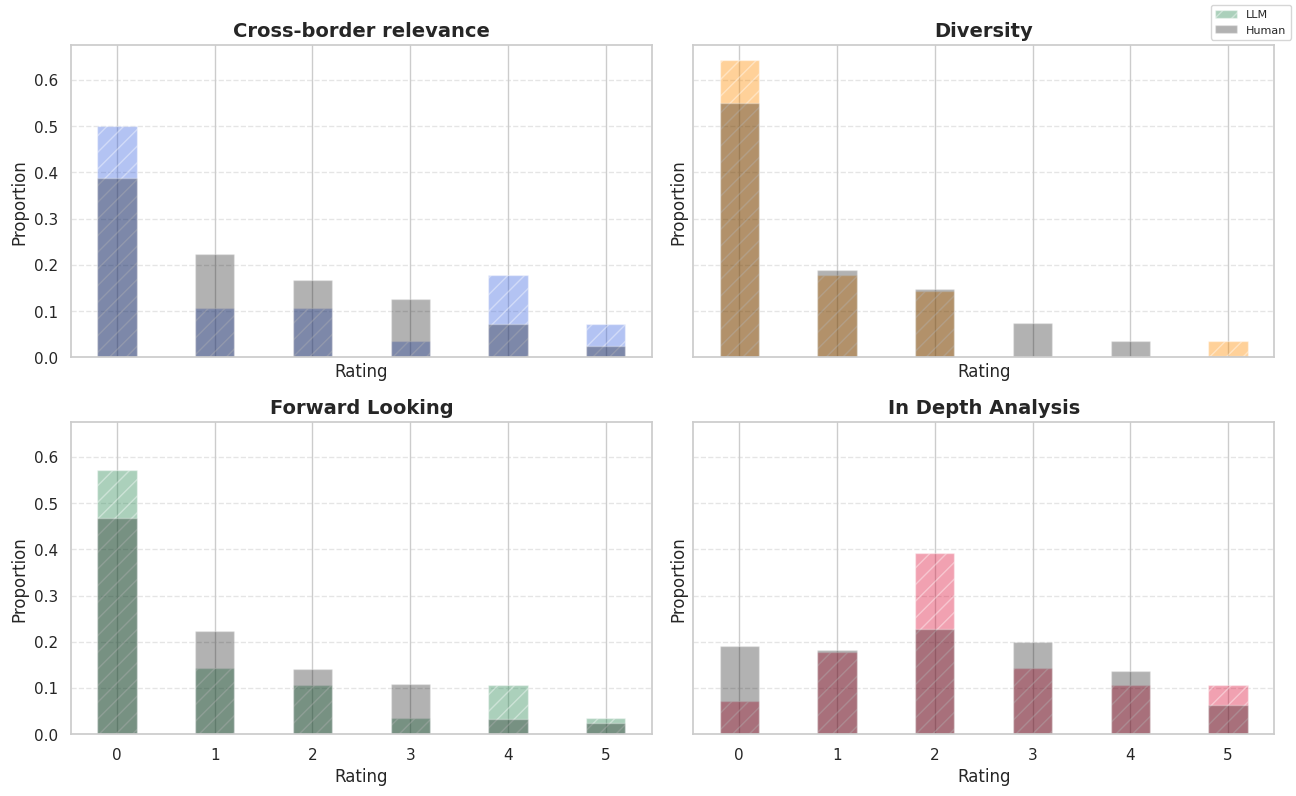}
        \caption{Llama 3.1 405B}
        \label{fig:dist_llama3_405b}
    \end{subfigure}

    \vspace{0.5cm}

    % --- Row 2 ---
    \begin{subfigure}[t]{0.47\textwidth}
        \centering
        \includegraphics[width=\linewidth]{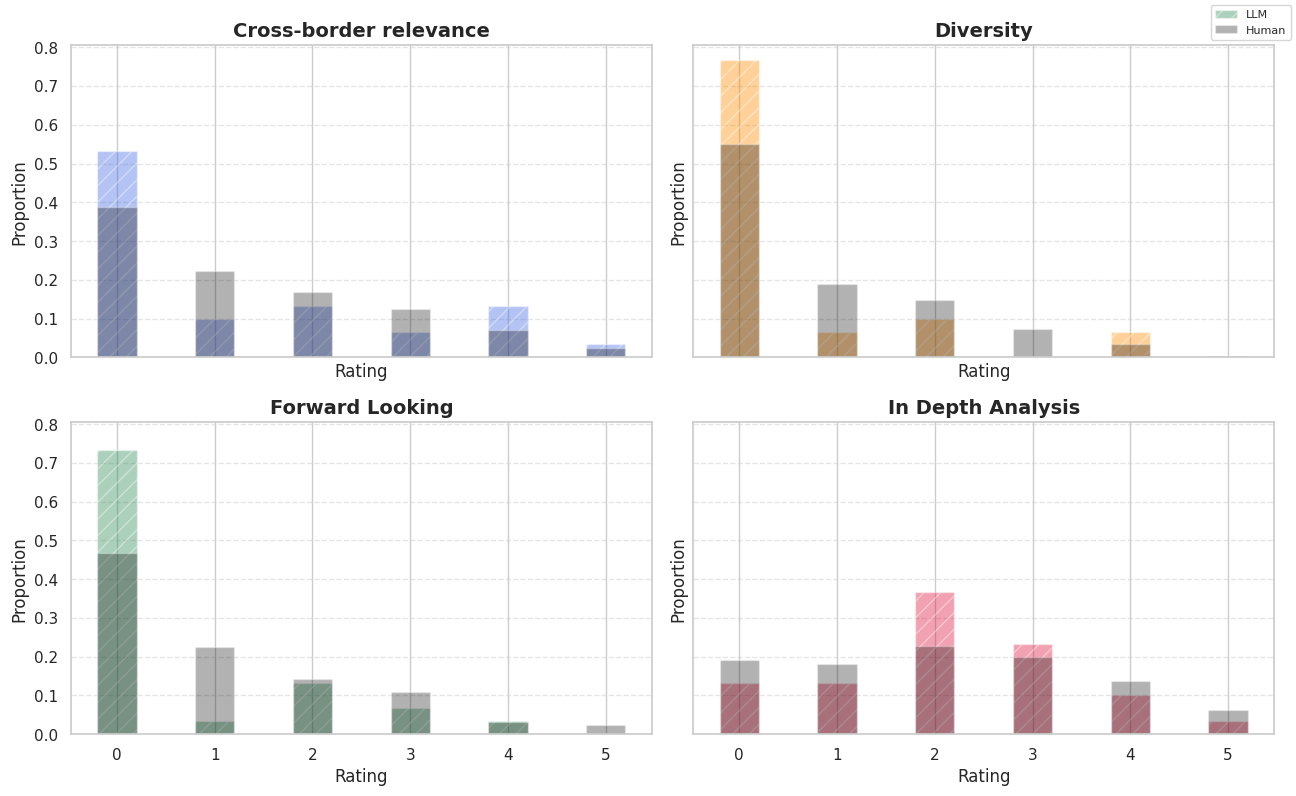}
        \caption{Llama 3 70B}
        \label{fig:dist_llama3_70b}
    \end{subfigure}
    \hfill
    \begin{subfigure}[t]{0.47\textwidth}
        \centering
        \includegraphics[width=\linewidth]{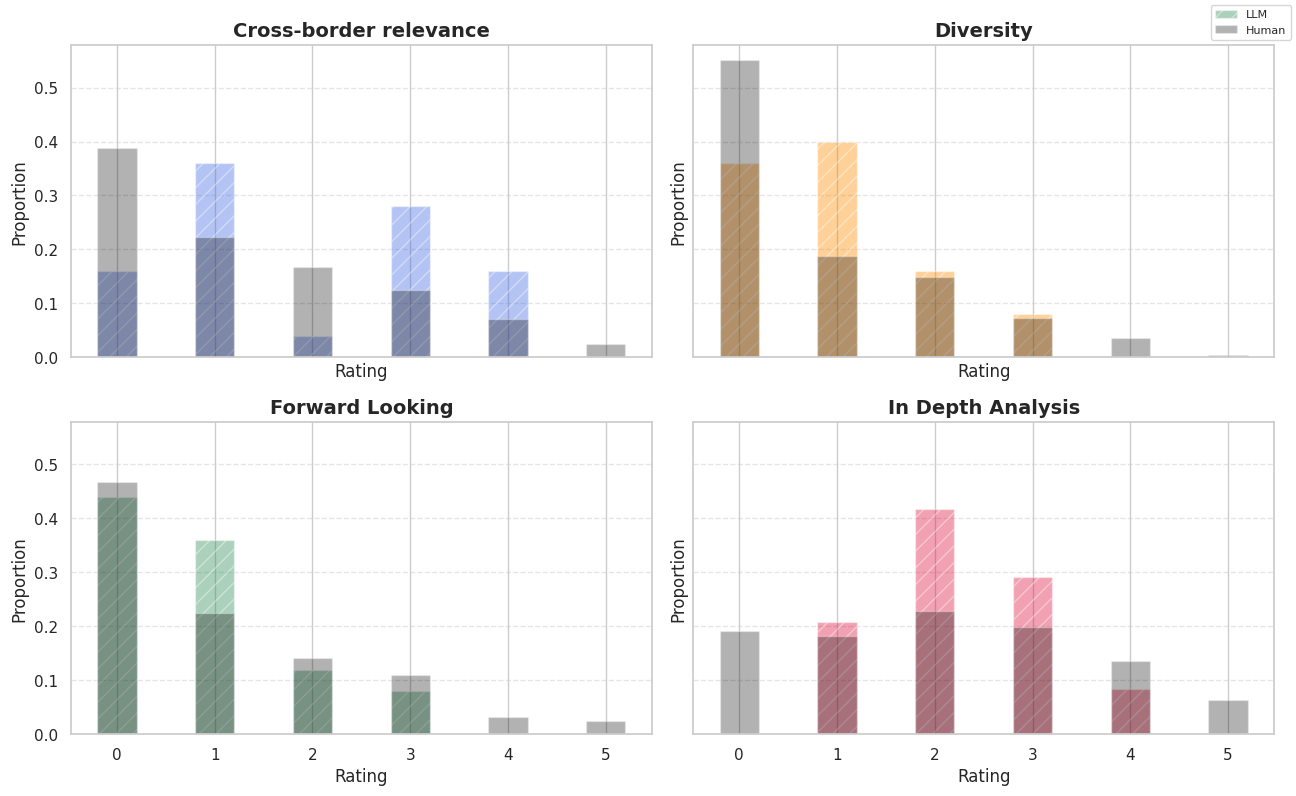}
        \caption{Mistral Large}
        \label{fig:dist_mistral_large}
    \end{subfigure}

    \vspace{0.5cm}

    % --- Row 3 ---
    \begin{subfigure}[t]{0.47\textwidth}
        \centering
        \includegraphics[width=\linewidth]{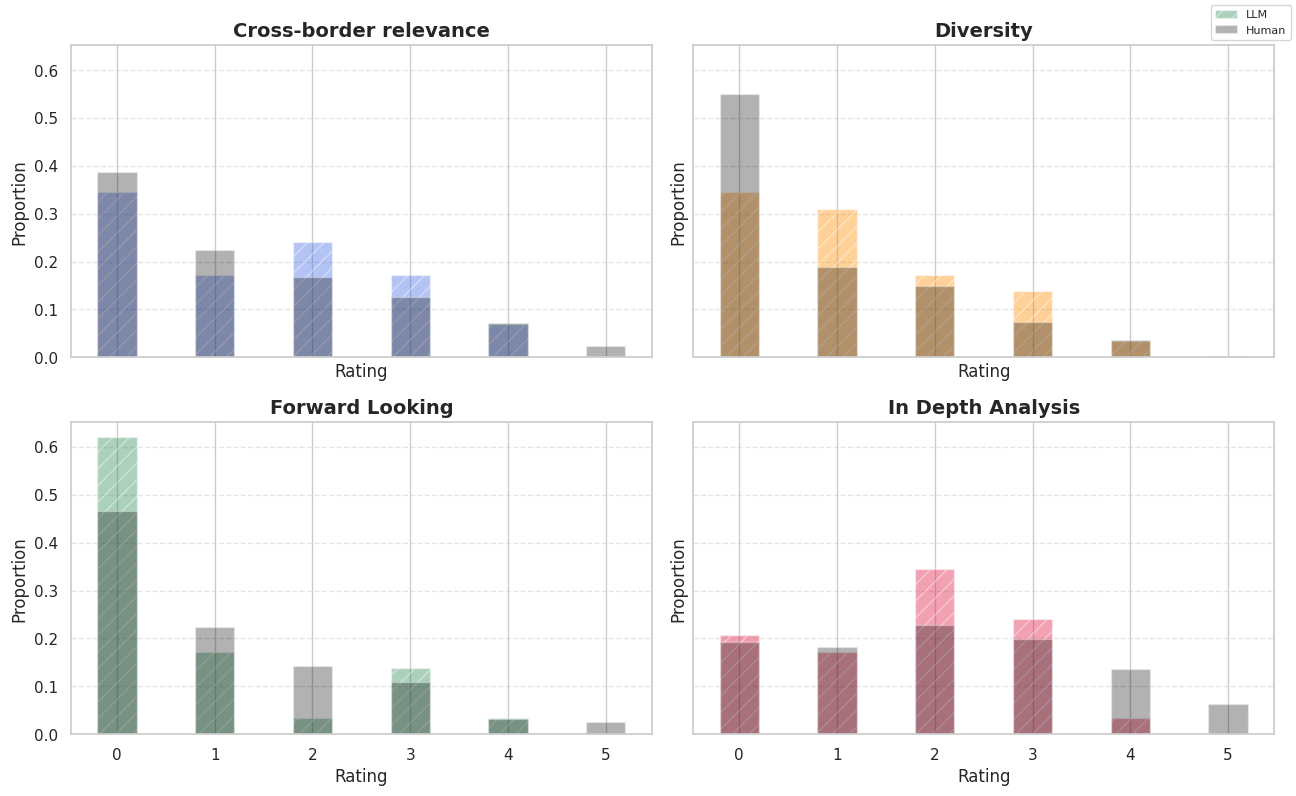}
        \caption{Mistral Nemo}
        \label{fig:dist_mistral_nemo}
    \end{subfigure}
    \hfill
    \begin{subfigure}[t]{0.47\textwidth}
        \centering
        \includegraphics[width=\linewidth]{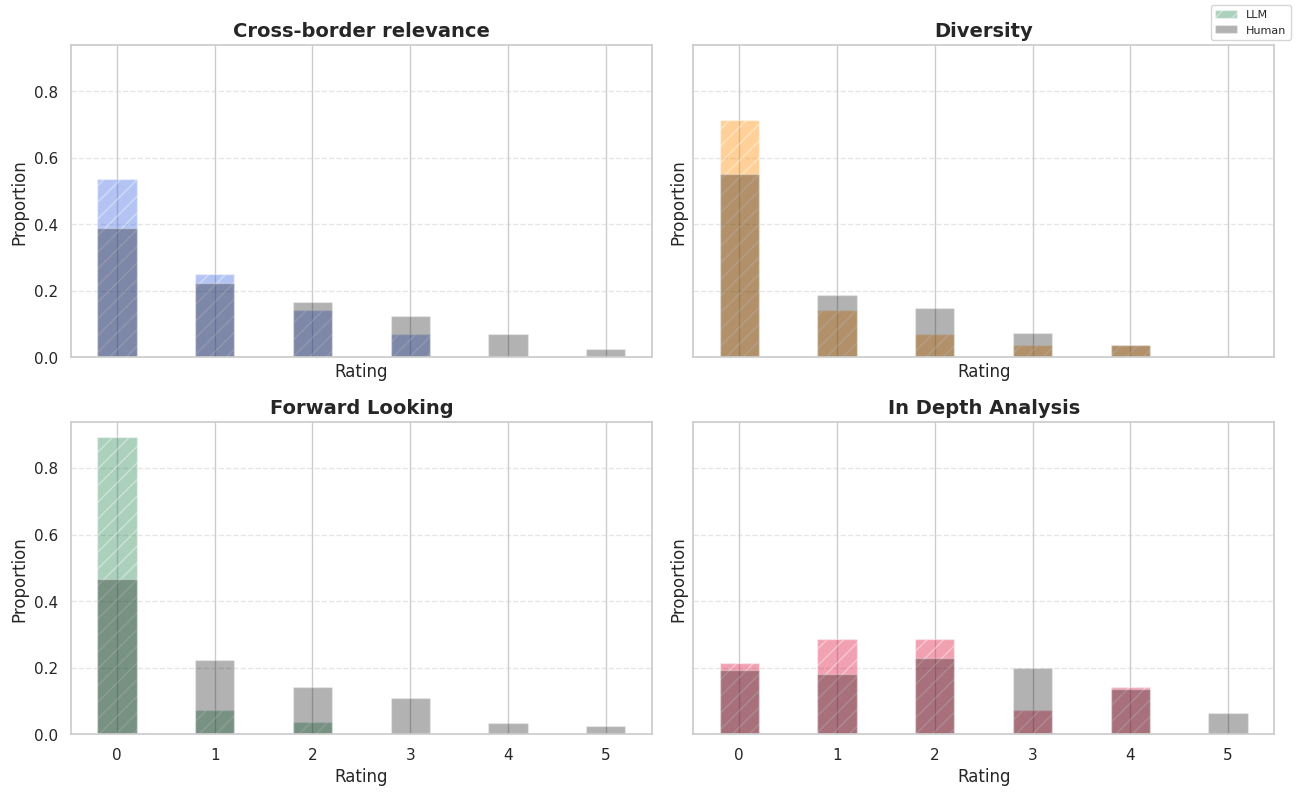}
        \caption{CommandR+}
        \label{fig:dist_commandr_plus}
    \end{subfigure}

    \vspace{0.5cm}

    % --- Row 4 ---
    \begin{subfigure}[t]{0.47\textwidth}
        \centering
        \includegraphics[width=\linewidth]{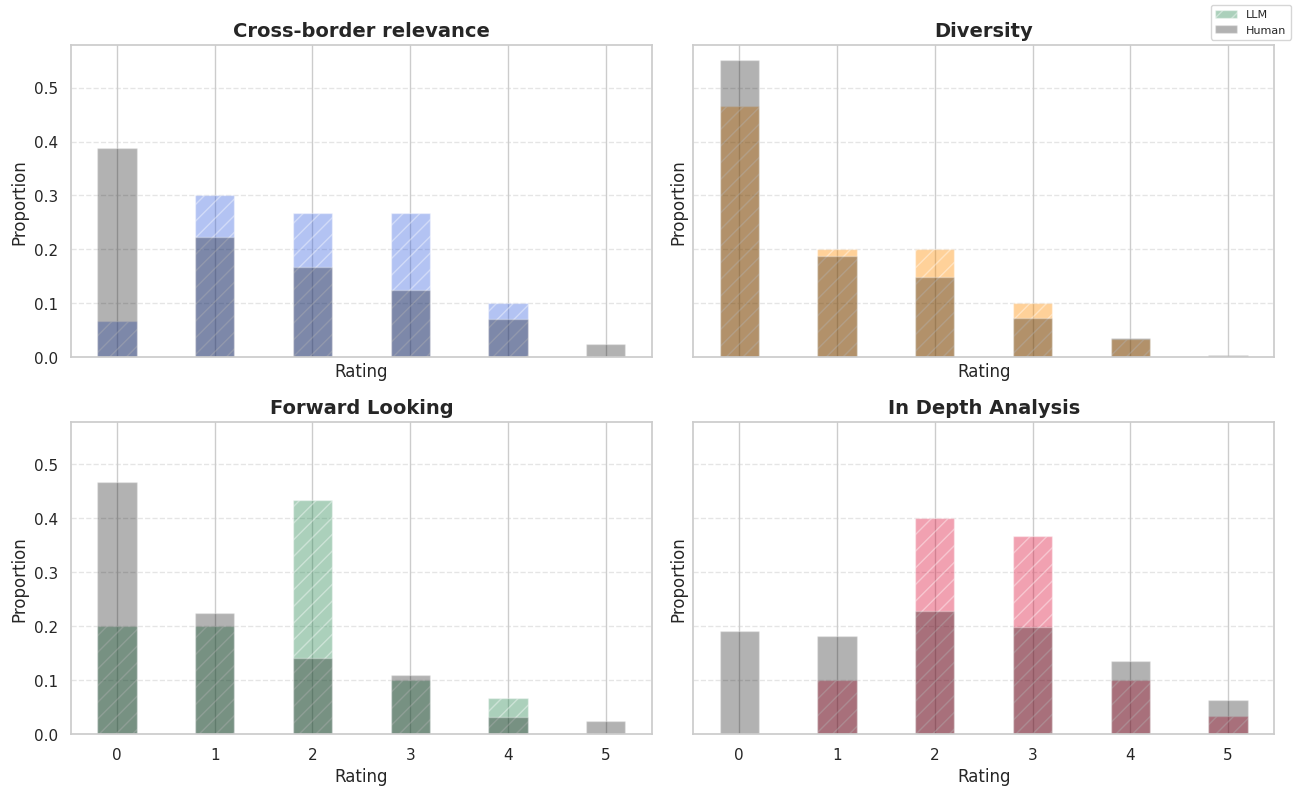}
        \caption{Qwen2.5 72B}
        \label{fig:dist_qwen2_72b}
    \end{subfigure}
    \hfill
    \begin{subfigure}[t]{0.47\textwidth}
        \centering
        \includegraphics[width=\linewidth]{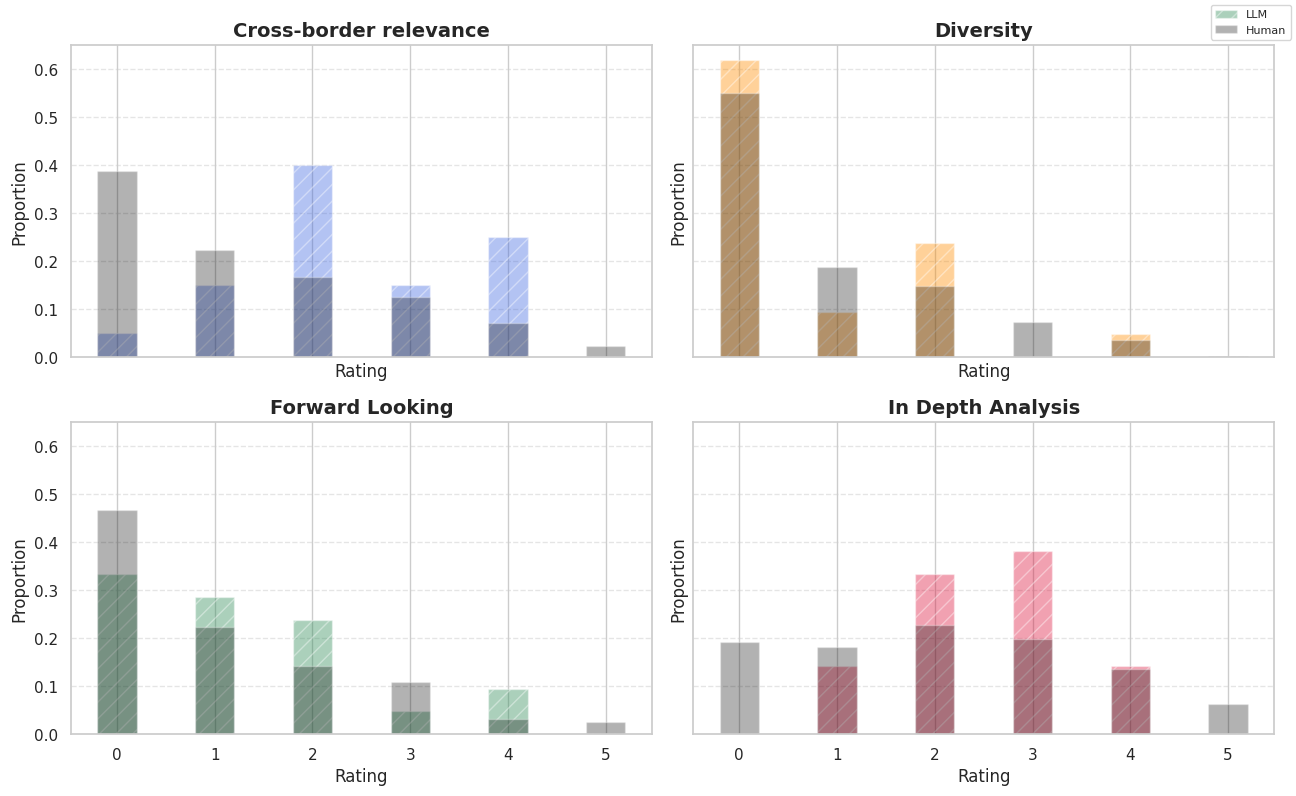}
        \caption{WizardLM 2 8x22B}
        \label{fig:dist_wizardlm2}
    \end{subfigure}

    \captionsetup{justification=centering}
    \caption{Rating distribution comparison: For each LLM, distributions of assigned scores per criterion are shown, overlaid with corresponding human editor distributions for direct comparison.}
    \label{fig:rating_distributions_comparison}
\end{figure}

To understand how individual LLMs' rating behaviors compare to that of human editors, we analyzed the normalized distribution of assigned scores for each criterion. This normalization allows for a direct comparison of the frequency with which different score values were used across the rating scale, irrespective of the absolute number of ratings, shown in figure~\ref{fig:rating_distributions_comparison}.

The comparative analysis reveals a promising picture of alignment and divergence. We observe that many LLMs exhibit a similar tendency for stringent scoring in "Diversity" and "Forward Looking." The modal score for these criteria is almost consistently 0 across both human and LLM raters. For "Diversity," the human proportion at 0 is 0.551. Several LLMs, such as Llama 3 and CommandR+, even exceed this human strictness. Similarly, for "Forward Looking," the human proportion is 0.467, and many LLMs demonstrate comparable or higher proportions, such as CommandR+ and Llama 3 70B. This indicates a shared alignment in the perception of these criteria. 

On the other hand, for "In Depth Analysis," both human and LLM distributions are notably more balanced, with lower proportions at 0 and higher concentrations in the mid-range ($1-3$). This suggests that "In Depth Analysis" is arguably the most straightforward criterion to assess for both human and LLMs.

Additionally, we notice variability and divergence for "Cross-border Relevance", as this criterion presents the most significant variations. While humans show a clear skew towards 0 (0.387), some LLMs are considerably more lenient or have a notably different distribution, whereas others are even more stringent, with Mistral Nemo exhibiting the most similarity in distribution.

\subsection{Average rating distribution per criteria}

\begin{figure}[htbp]
    \centering
    \begin{subfigure}[t]{0.8\textwidth}
        \centering
        \includegraphics[width=\linewidth]{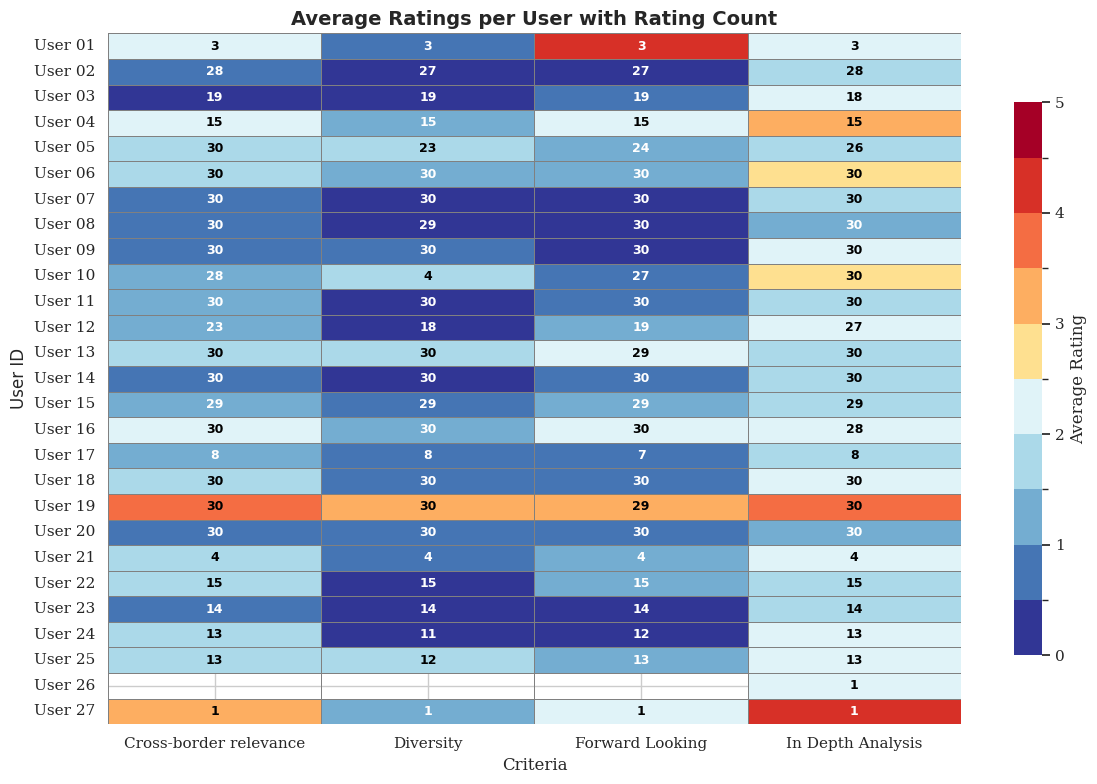}
        \caption{Human editors' average ratings across all articles per criteria. }
        \label{fig:avgeditors}
    \end{subfigure}

    \vspace{0.5cm}

    \begin{subfigure}[t]{0.8\textwidth}
        \centering
            \includegraphics[width=\linewidth]{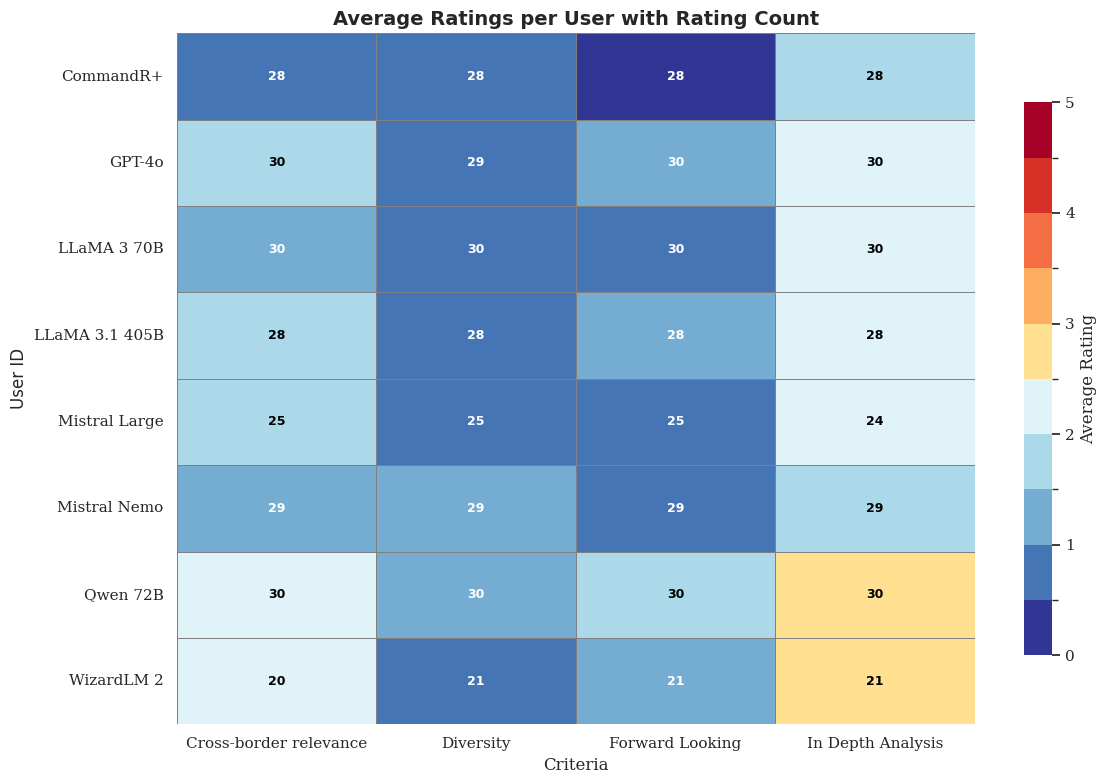}
        \caption{LLMs' average ratings aggregated per model for per criteria.}
        \label{fig:avg_llm}
    \end{subfigure}

    \captionsetup{justification=centering}
\caption{Average ratings for all articles per criteria for both LLMs and Human raters.}
    \label{fig:avg_rating_dist}
\end{figure}

To assess individual rater behavior and the overall participation landscape, we analyzed the average rating assigned by each human editor for each criterion, alongside the number of articles they rated. This detailed breakdown, presented in figure~\ref{fig:avgeditors}, showcases individual scoring tendencies and the uneven contribution patterns within the human expert panel.

The data on participation reveals a substantial range in the number of articles rated by individual editors. While a core group of editors (e.g., User 06, User 14, User 18) provided comprehensive coverage, rating nearly all 30 articles across all criteria, a significant portion of the panel contributed to a much lesser extent (e.g., User 01 rated only 3 articles, User 21 rated 4, and User 26 provided only 1 rating across all criteria). This variability in participation underscores the practical challenges inherent in large-scale human annotation projects and indicates that some articles or criteria received considerably fewer expert judgments than others, potentially affecting the density and representativeness of the human ground truth.

Correspondingly, the analysis of average ratings provides further quantitative evidence for the inter-rater variability. Editors exhibited distinct individual scoring tendencies:

\begin{itemize}
    \item \textbf{"Strict" Raters:} A number of editors, exemplified by User 02, User 03, and User 08, consistently assigned significantly lower average scores across all criteria. For instance, User 02's average scores ranged from 0.18 (Diversity) to 1.89 (In-Depth Analysis), indicating a stringent interpretation of the editorial guidelines or a higher threshold for an article to meet the criteria.
    \item \textbf{"Lenient" Raters:} Conversely, User 19 stands out as a particularly "lenient" rater, with average scores consistently above 3 across most criteria (e.g., 3.7 for Cross-border relevance and 3.2 for Diversity). Other editors, such as User 04 and User 16, also demonstrated a tendency towards higher average ratings relative to the panel's overall mean.
\end{itemize}

Furthermore, to identify potential anomalies in human annotations, we applied the standard Interquartile Range (IQR) method for outlier detection. For each article and criterion, we calculated the first quartile ($Q_1$) and third quartile ($Q_3$) of the rating distribution across raters. An individual rating was considered an outlier if it fell outside the range:

\[
[Q_1 - 1.5 \times \text{IQR}, \; Q_3 + 1.5 \times \text{IQR}]
\quad \text{where} \quad \text{IQR} = Q_3 - Q_1
\]

\begin{figure}[htbp]
    \centering
    \includegraphics[width=0.5\textwidth]{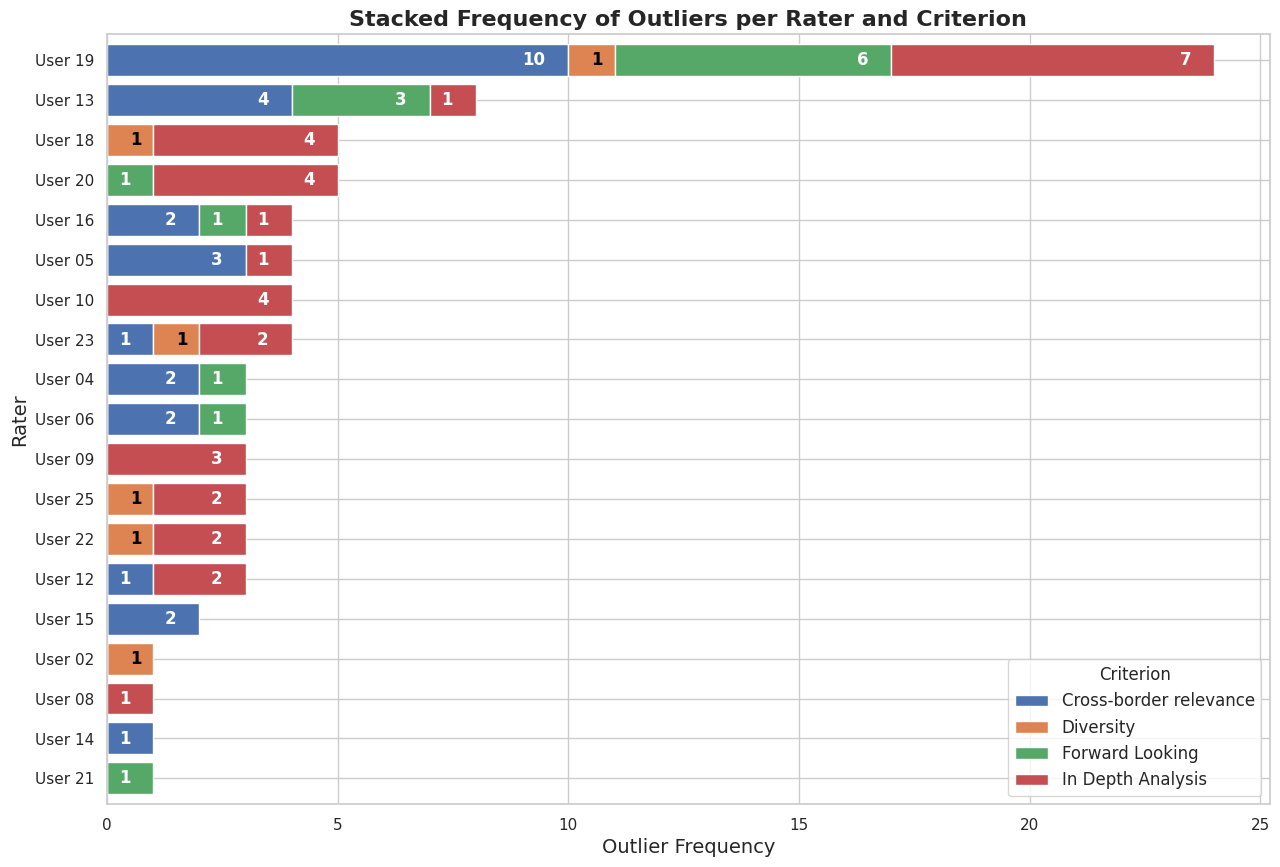}
    \captionsetup{justification=centering}
    \caption{Human outlier frequency per criteria.}
    \label{fig:rater_outlier}
\end{figure}

This approach helps flag unusually high or low scores, and figure~\ref{fig:rater_outlier} presents the frequency of such outliers per criterion.

These divergent average rating behaviors directly contribute to the overall spread and variability observed in the human ratings. Identifying these individual biases is crucial for understanding the composition of the "ground truth": it represents an aggregated consensus amidst varied individual interpretations, rather than a monolithic agreement. Such insights are valuable for developing future rater calibration strategies in similar annotation projects.

On the other hand, figure~\ref{fig:avg_llm} shows the average rating per LLM per criteria, visualizing the differences in LLM ratings. Qwen (average In-Depth Analysis: 2.57) and WizardLM 2 (average In-Depth Analysis: 2.52) are relatively more generous for 'In-Depth Analysis' compared to other models. Conversely, CommandR+ exhibits a generally strict rating tendency across most criteria, with average scores such as 0.14 for 'Forward Looking' and 0.54 for 'Diversity'. The overall pattern of LLM ratings across criteria is similar to that of the human consensus, where 'In-Depth Analysis' and 'Cross-border relevance' generally receive higher average ratings compared to 'Diversity' and 'Forward-Looking'.

\subsection{Analysis of Rating Distributions by Criterion}

Figure~\ref{fig:rater_agreement_visuals} presents rating distributions for each article across four evaluation criteria: \textit{Cross-border relevance}, \textit{Diversity}, \textit{Forward Looking}, and \textit{In Depth Analysis}. Human annotator ratings are shown as boxplots, while LLM predictions are overlaid as individual colored markers. Vertical dashed lines indicate the mean and median of human scores for each article. Below, we analyze the patterns observed for each criterion. 

\subsubsection{Cross-border Relevance}

Human ratings for cross-border relevance show strong agreement across most articles, with narrow interquartile ranges and minimal outliers. This is reflected in the high inter-rater reliability score (ICC = 0.891). LLM predictions are generally aligned with the central tendency of human judgments, though some models, such as WizardLM2, occasionally assign higher-than-average scores. Overall, LLMs perform relatively well on this criteria.

\subsubsection{Diversity}

Diversity exhibits greater variation among human raters, both in terms of boxplot spread and outlier presence. The ICC score of 0.791 suggests moderate agreement. LLMs tend to be more conservative, often rating articles lower than human annotators on this dimension. This may indicate difficulty in capturing implicit indicators of diversity, such as inclusion of underrepresented voices or geographic spread, without explicit cues in the text.

\subsubsection{Forward Looking}

Forward-looking evaluations have the lowest inter-rater reliability (ICC = 0.747), highlighting the subjective nature of this criterion. While some articles, typically those concerning emerging trends or long-term impacts, receive consistent human scores, many others produce wide disagreement. LLM predictions also vary, with some models consistently underestimating this dimension. This suggests challenges in detecting speculative or future-oriented framing unless overtly stated.

\subsubsection{In Depth Analysis}

This criterion yields the highest inter-rater agreement (ICC = 0.942), as reflected by the tight clustering of boxplots. Human raters agree strongly on which articles exhibit analytical rigor. LLMs perform well on this trait, closely mirroring human assessments, especially for long-form or investigative articles. Occasional overestimation is observed for shorter or structurally simple articles, but overall alignment is high.

\subsection{Comparative Ranking Analysis: Human vs. LLM}

As shown previously in figure~\ref{fig:rater_agreement_visuals}, LLMs align more closely with human ratings on well-defined, structural traits such as \textit{In Depth Analysis} and \textit{Cross-border relevance}. In contrast, they show slightly greater divergence on more abstract or subjective traits like \textit{Diversity} and \textit{Forward Looking}, despite refining the guide with incremental guide~\ref{annex:psa_crit}. These findings suggest that future model improvement should focus on enhancing contextual sensitivity and definitional clarity for complex criteria. While LLMs can aid in evaluation workflows, human oversight remains essential for nuanced judgment.

A key objective was to determine if an LLM could capture the collective editorial signal or "group wisdom." To test this, we performed a comparative ranking analysis. First, we calculated the average score for each article based on the aggregated human ratings. These articles were then ranked to create a human-consensus "top list" in Table~\ref{tab:top5_model_selections}. This list was compared against the ranking produced by the LLM for the same number of articles.

The analysis revealed a strong and promising alignment at the top of the rankings. When comparing the list of the top 5 articles identified by the human panel (using their averaged scores) with the top 5 articles identified by the LLM, we observed a 75\% overlap in the best models like Qwen and Mistral Nemo.

This high degree of concordance in a practical, top-k selection scenario is a significant finding. It suggests that while individual human scores may exhibit variability, the LLM is effectively able to approximate the collective judgment and identify content that, on average, aligns with the highest value assessments of the expert group. This demonstrates the LLM's strong potential to automate the identification of premium content with a level of accuracy that reflects the collective judgment of a team of editors.

\begin{table}[htbp]
\centering
\caption{Top-5 article selections per model. \textbf{Bold values} indicate articles absent from the human top-5 ranking.}
\label{tab:top5_model_selections}
\begin{tabular}{l | l l l l l}
\toprule
\textbf{Model} & \textbf{1} & \textbf{2} & \textbf{3} & \textbf{4} & \textbf{5} \\
\midrule
Human         & Article 05 & Article 24 & Article 12 & Article 03 & Article 01 \\
CommandR+     & Article 05 & Article 24 & \textbf{Article 19} & \textbf{Article 21} & \textbf{Article 18} \\
LLaMA 3 70B   & Article 03 & Article 24 & \textbf{Article 28} & \textbf{Article 04} & Article 05 \\
Mistral Large & \textbf{Article 25} & \textbf{Article 04} & \textbf{Article 28} & Article 24 & Article 03 \\
Mistral Nemo  & Article 03 & Article 05 & Article 12 & Article 24 & \textbf{Article 28} \\
LLaMA 3.1 405B& Article 05 & \textbf{Article 28} & Article 24 & Article 03 & \textbf{Article 29} \\
Qwen 72B      & Article 24 & Article 03 & Article 12 & Article 05 & \textbf{Article 29} \\
GPT-4o        & Article 05 & Article 03 & Article 24 & \textbf{Article 19} & \textbf{Article 28} \\
WizardLM 2    & Article 03 & \textbf{Article 08} & \textbf{Article 28} & \textbf{Article 04} & Article 01 \\
\bottomrule
\end{tabular}
\end{table}

\begin{table}[htbp]
\centering
\caption{Model performance based on NDCG@5 and Precision@5, computed from top-5 article selections out of 30 total articles. Higher values indicate stronger alignment with human top-5 rankings.}
\label{tab:ndcg_precision_scores}
\begin{tabular}{l r r}
\toprule
\textbf{Model} & \textbf{NDCG@5} & \textbf{Precision@5} \\
\midrule
Qwen 72B          & 0.9495 & 0.75 \\
Mistral Nemo      & 0.9421 & 0.75 \\
GPT-4o            & 0.9332 & 0.60 \\
LLaMA 3.1 405B    & 0.9065 & 0.60 \\
CommandR+         & 0.8168 & 0.40 \\
LLaMA 3 70B       & 0.8122 & 0.60 \\
WizardLM 2        & 0.6551 & 0.40 \\
Mistral Large     & 0.5116 & 0.40 \\
\bottomrule
\end{tabular}
\end{table}

\subsection{Challenges and Limitations}
\label{sec:lim}
This study provides an initial examination of LLMs for value-driven news curation, but it operates under several constraints. These limitations define the scope of the current findings and inform future research directions.

First, the dataset size limited the breadth of our findings. The study utilized 30 news articles. While selected for representativeness within the "A European Perspective" project, this quantity restricts the generalizability of results across the full spectrum of news content and topics. A larger and more diverse corpus of articles would offer greater statistical confidence in the observed patterns of alignment between LLMs and human judgments.

Second, the human editor participation rate affected the completeness of the ground truth data. Despite engaging 30 experienced editors, the overall completion rate for article ratings was 68.9\%. This resulted in a sparse rating matrix, where some articles or specific criteria received fewer human judgments than others. This uneven contribution necessitated adjustments during the calculation of inter-rater reliability (ICC). Specifically, to compute ICC, some articles or individual rater contributions had to be omitted from certain calculations due to missing data. This impacts the direct interpretability of the reported reliability metrics and the density of the human-annotated baseline. The aggregated human "ground truth" therefore represents a consensus among varied individual contributions rather than a uniformly dense set of judgments.

Third, the subjectivity inherent in defining and applying PSM criteria contributes to data variability. Even with refined guidelines, human editors exhibited a range of interpretations for concepts like "Forward Looking" or "Diversity." While our inter-rater reliability analysis quantifies this variability, it underscores the challenge of establishing a perfectly consistent human ground truth for abstract values. This variability in human annotation has implications for the maximum achievable alignment with LLMs.

Fourth, the translation of all news articles into English before LLM processing introduces a potential compromise. Original news content often contains linguistic, cultural, and political nuances specific to its source language. Performing analysis on translated text may lead to the loss of subtle elements relevant to PSM criteria. This could affect the LLM's capacity to assess an article's merit fully, particularly for criteria like "Cross-border relevance" or "Diversity" which might be tied to specific local contexts.

Finally, while LLMs offer mechanisms for transparency through rationale generation~\cite{bommasani2021opportunities}, the inherent non-deterministic nature of these models means that identical prompts may not always yield identical responses~\cite{blodgett2020language}. This necessitates careful experimental design to account for internal consistency when evaluating LLM performance. The internal logic and potential biases embedded within LLM training data also remain external to our direct analysis, representing a deeper layer of potential limitation for large-scale, ethical deployment~\cite{buyl2024large}.

\section{Conclusion and Future Work}

% For the future, we want to expand this for a bigger selection, bigger rating. 
% We want to use the insights from reason the llms are giving to improve the prompts of the LLMs (positive/negative prompts). 
% Incorporate readers signal, and see if readers respond higher to articles that are highly rated and ranks. 

% This study demonstrates the potential for Large Language Models to assist in scaling content curation based on editorial values. The observed alignment between LLM ratings and human editorial judgments, particularly for top-ranked content show promising results. This work establishes a foundation for automated, value-driven systems in the news ecosystem.
In this study we successfully demonstrates the capacity of Large Language Models to assist in content curation based on predefined editorial values.
Our empirical comparison of LLM-generated ratings with expert human editorial judgments revealed a notable alignment, particularly in identifying articles that consistently ranked high across multiple PSM criteria. This observed convergence underscores the potential of LLMs to effectively augment and, in part, automate the highly nuanced task of content evaluation previously exclusive to human editors. This capability directly addresses the pressing challenge of scalability in traditional editorial processes, offering a pathway for efficiently processing vast volumes of news content.

Future research will extend this work by significantly expanding data collection for both articles and human ratings, thereby building a more comprehensive ground truth. We will analyze LLM-generated rationales to refine prompt engineering, aiming for improved model accuracy. A key objective involves integrating reader signals to assess audience response to PSA-rated content and exploring PSA-based thresholding for recommendation systems. 

\section*{Acknowledgments}
This project has been co-financed by the European Union's Preparatory Action - "European Media Platforms", in the framework of the LC-02953453, DNP EUROPE project.
We want to extend our thanks to the AEP editorial team and all the AEP associated editors from the different participating organizations that contributed to rating the articles.

\bibliography{references}

\appendix
\section*{Appendix: Detailed PSA Criteria}
\label{annex:psa_crit}

In this section we present the separate guides for both the human editors and for the LLMs.  
Note that the incremental approach is used for both to guide the ratings. 

\subsection*{In-depth Analysis}

\subsubsection*{Editors’ Guide}

Current affairs stories that orientate and inform audiences about important issues or events by providing 
analysis and context, supported by verifiable facts and/or data. Stories are ranked higher that analyse 
lead stories in current news reporting, with essential 'what you need to know' details. The source of the 
analysis can either be a journalist (i.e., the article’s author) specialising in the subject, or researchers 
interviewed for the story who have a profound knowledge of the subject. The story length per se is not a 
factor in determining the depth of analysis provided but rather if the content conveys new and broader 
insights on the issue.

\textbf{Short definition:} Stories that orientate audiences about important issues or events by providing expert analysis based on facts.

\vspace{1em}
\subsubsection*{LLM Guide}

Current affairs stories that orientate and inform audiences about important issues or events by providing 
analysis and context, supported by verifiable facts and/or data. The target audience are all Europeans 
seeking independent, impartial and accurate public service journalism as defined by the EBU public 
service values and editorial principles and media users interested in key tenets as defined by 
representative bodies like the Council of Europe especially how these tenets impact their lives as citizens. Stories are ranked higher that analyse lead stories in current news reporting, with essential 
'what you need to know' details (does include background information). The expert voice providing the 
analysis can either be a journalist (i.e., the article’s author does include a byline. Bylines mentioning only 
agencies such as EFE, keystone-SDA, Associated Press, Reuters are excluded) specialising in the 
subject, or specialized researcher/professional interviewed for the story by the reporting media (does 
include name with their title or expertise and/or institution they represent) who have a profound 
knowledge of the subject. The story length per se is not a factor in determining the depth of analysis 
provided but rather if the content conveys broader insights on the issue, such as the impact on the 
parties involved, whether individuals, associations, institutions, organisations or companies, and/or 
provides detailed answers to how and why questions for additional context.

\vspace{1em}
\subsubsection*{Incremental Approach (Scoring Guidance)}

\begin{description}
  \item[] +1 provides verifiable facts. For example, sources are provided for factual statements, such as: “Spain's aquifers are at an all-time low, according to Spain's environment ministry”.
  \item[] +1 provides need-to-know or essential details
  \item[] +1 includes an expert voice as defined above (specialised journalist/researcher)
  \item[] +1 includes a minimum of two sources. All sources are mentioned by name
  \item[] +1 provides broader insights
\end{description}

\subsection*{Diversity}

\subsubsection*{Editors’ Guide}

Stories on under-reported communities or groups. Stories are ranked higher which promote voices, 
views, and personal testimonies from across Europe's social landscape, including those of vulnerable or 
marginalised groups and communities.

\textbf{Short definition:} Stories about under-reported communities that add new voices and perspectives.

\vspace{1em}
\subsubsection*{LLM Guide}

Stories on under-reported communities or groups (does include national or ethnic minorities and/or 
linguistic minorities). Stories promote citizen voices, views, and personal testimonies (does include a 
person or person’s opinion—other than the author, view or perspective—in a direct or indirect quote) from 
across Europe's social landscape, including those of vulnerable or marginalised groups and communities 
(does include specific mentions of how or why the group or groups are disadvantaged, due to age, 
gender, sexual orientation, race, ethnicity, or disability). Ethnic minorities can be divided into two 
groups: indigenous minorities that may or may not overlap borders, such as the Roma, Sami, or Basques, 
and minorities with migrant backgrounds, such as ethnic Turks, North and West Africans, and Russians. 
The migrant communities can also be considered linguistic minorities like Russians in the Baltic states. 
For this exercise, the Italian-speaking region of Switzerland, the German-speaking region in Belgium, or 
the Sami-speaking region in Sweden and others in similar circumstances qualify as a linguistic minority.

\vspace{1em}
\subsubsection*{Incremental Approach (Scoring Guidance)}

\begin{description}
  \item[] +1 focus is on an ethnic or linguistic minority
  \item[] +1 focus is on a disadvantaged age group or on a disability
  \item[] +1 focus is on a gender and/or sexual minority
  \item[] +1 includes personal view or testimony
  \item[] +1 provides details on vulnerability and/or explains the impact on under-reported community
\end{description}

\subsection*{Cross-border Relevance}

\subsubsection*{Editors’ Guide}

Stories that have a universal, cross-border relevance that matter to all citizens in Europe even if the 
report is about an event or issue of a local or regional community. The content, which can include news 
stories, should assist the public in understanding how social and political challenges like and including 
climate change, migration, health crises and violence conflicts in border regions are addressed and 
overcome in comparable environments. Stories are ranked higher if they offer original reporting and 
address issues of European values (based on Council of Europe's values: Human Rights, Democracy, 
Rule of Law) and identity. Stories are also considered that analyse the impact on citizens of policies, 
decisions, and actions of European institutions.

\textbf{Short definition:} Stories that have a universal aspect resonating with a broader European audience.

\vspace{1em}
\subsubsection*{LLM Guide}

Stories that have a universal, cross-border relevance that matter to all citizens in Europe (defined by the 
EBU public service values and editorial principles and representative bodies like the Council of Europe) 
even if the report is about an event or issue of a local or regional community (does provide sufficient 
details and/or context for a wider European audience). The content, which can include news stories, 
should assist the public in understanding how social and political challenges like and including climate 
change, migration, health crises and violence conflicts in border regions are addressed and overcome in 
comparable environments. Stories receive an additional point if they offer original reporting by a 
journalist (i.e., the article’s author does include a byline. Bylines mentioning only agencies such as EFE, 
keystone-SDA, Associated Press, Reuters are excluded) and address issues of values important to all 
Europeans (Human Rights, Democracy, Rule of Law) and identity. Stories are also considered that 
analyse the impact on citizens of policies, decisions, and actions of European institutions (does include a 
specific mention by name or acronym of the decision or action as well as the issuing institution).

\vspace{1em}
\subsubsection*{Incremental Approach (Scoring Guidance)}

\begin{description}
  \item[] +1 includes at least one of the listed values
  \item[] +1 offers detailed explanation of social or political challenge
  \item[] +1 focuses on a local/regional/national response to global issues (i.e. climate change, violent conflicts in border regions, migration, public health…)
  \item[] +1 focuses on the broader impact on people across Europe
  \item[] +1 includes original reporting
\end{description}

\subsection*{Forward Looking}

\subsubsection*{Editors’ Guide}

Stories that are forward-looking and inspiring, showcasing new or constructive approaches to problems of 
a universal nature. Stories are ranked higher that offer details of innovative responses to major 
challenges of our time (climate change, migration, inflation etc). Also, stories highlighting ways 
to improve people’s everyday lives can be considered, since these, in part, help counter ‘news avoidance’. 
Stories that focus on ideas such as proposals from politicians that lack detail, should be excluded.

\textbf{Short definition:} Stories showcasing innovative, forward-looking and constructive approaches to universal problems.

\vspace{1em}
\subsubsection*{LLM Guide}

Stories that are forward-looking (does include details of how parties/actors will engage on an issue), 
showcasing new or constructive approaches to problems of a universal nature (does include specific 
details of a constructive problem-solving approach). Stories receive an additional point that offer details of 
innovative responses to major challenges of our time (i.e. climate change, public health, migration, 
inflation, social conflict, AI, disinformation etc). Also, stories highlighting ways to improve people’s 
everyday lives (does include at least one aspect from the CoE list such as employment, housing, health, 
education, welfare, climate change mitigation) are considered, since these, in part, help counter ‘news 
avoidance’ by providing new perspectives and therefore lead to broader comprehension (exclude stories 
that are considered negative or repetitive). Stories that focus on ideas such as proposals from 
politicians that lack detail, should be excluded.

\vspace{1em}
\subsubsection*{Incremental Approach (Scoring Guidance)}

\begin{description}
  \item[] +1 explains how party/actor will engage on issue
  \item[] +1 offers details of at least one constructive approach
  \item[] +1 offers details of an innovative response to at least one of the following: climate change, biodiversity, public health, migration, inflation, social conflict, AI, disinformation
  \item[] +1 provides at least one way to improve people’s everyday lives
  \item[] +1 offers background information sufficient to contextualise issue or issues described
\end{description}
\section*{Dataset description}
The dataset contains all the anonymized rated articles by human editors and LLMs, as well as the aggregated traffic data on the articles rated by PSA that are featured on \url{https://yepnews.eu}.  
\end{document}